\documentclass[superscriptaddress,letterpaper]{revtex4}

\usepackage{graphicx}
\newcommand{\MW}{M_W}
\newcommand{\MZ}{M_Z}
\newcommand{\sweff}{\sin^2\theta_{\mathrm{eff}}}
\newcommand{\gammal}{\Gamma_l}
\newcommand{\mt}{m_t}
\newcommand{\MH}{M_H}

\newcommand{\De}{\Delta}
\newcommand{\de}{\delta}
\newcommand{\msbar}{$\overline{\rm MS}$}
\def\order#1{${\cal O}(#1)$}
\def\ul#1{\underline{#1}}
\newcommand{\al}{\alpha}
\newcommand{\als}{\alpha_s}
\newcommand{\ifb}{${\mathrm{fb}}^{-1}$}

\begin{document}

\thispagestyle{empty}
\setcounter{page}{0}
\def\thefootnote{\fnsymbol{footnote}}


\begin{flushright}
BNL--HET--02/5\hfill
FT2002-01\\
DCPT/02/14\hfill
IPPP/02/07\\
UB-HET--02--02\hfill
PITHA 02/04\\
FERMILAB-Conf-02/010-T\hfill
KA-TP-3-2002\\
hep-ph/0202001\\
\end{flushright}

\vspace{1cm}

\begin{center}

{\Large\sc {\bf Present and Future Electroweak Precision
Measurements and the Indirect Determination of the Mass of the Higgs
Boson$
^{\ddag}$}}

\vspace{1cm}

{\sc {\bf The Snowmass Working Group on Precision Electroweak
Measurements\\}
U.~Baur$^{\,1}$%
,
R.~Clare$^{\,2,^*}$%
,
A.~Denner$^{\,3}$%
,
J.L.~Diaz~Cruz$^{\,4}$%
,
S.~Dittmaier$^{\,5}$%
,
J.~Erler$^{\,6,^*}$%
,
M.~Gr\"unewald$^{\,7}$%
,
S.~Heinemeyer$^{\,8,^*}$%
,
U.~Heintz$^{\,9}$%
,
M.~Kraemer$^{\,10}$%
,
H.E.~Logan$^{\,11}$%
,
K.~M{\"o}nig$^{\,12}$%
,
M.~Narain$^{\,9}$%
,
M.~Roth$^{\,13}$%
,
M.~Schmitt$^{\,14}$%
,
D.~Wackeroth$^{\,15}$%
,
G.~Weiglein$^{\,16}$%
,
D.R.~Wood$^{\,17,^*}$%
 and
J.~Wudka$^{\,2}$%
}

{\small

\vspace*{0.5cm}
$^1${State University of New York at Buffalo, Buffalo, NY 14260,
USA}

$^2${University of California, Riverside, CA 92521, USA}

$^{3}${Paul Scherrer Institut, Villigen PSI, Switzerland}

$^4${IPUAP-BUAP, 72570 Puebla, Pue., Mexico}

$^{5}${DESY, Hamburg, Germany}

$^6${Dept.\ of Physics and Astronomy, University of Pennsylvania,
PA 19146, USA and  Universidad Nacional Autonoma de Mexico,Instituto de
Fisica, 01000 Mexico D.F., Mexico}

$^7${RWTH Aachen, Aachen, Germany}

$^8${HET Physics Dept., Brookhaven Natl.\ Lab., NY 11973, USA}

$^9${Boston University, Boston, MA 02215, USA}

$^{10}${University of Edingburgh, Edingburgh, UK}

$^{11}${Fermilab, PO Box 500, Batavia, IL 60510-0500, USA}

$^{12}${DESY, Zeuthen, Germany}

$^{13}${Universitaet Karlsruhe, Karlsruhe, Germany}

$^{14}${Northwestern University, Evanston, IL, USA}

$^{15}${University of Rochester, Rochester, NY 14627, USA}

$^{16}${Institute for Particle Physics Phenomenology, Durham, UK}

$^{17}${Northeastern University, Boston, MA 02115, USA}

}

\end{center}

\vspace*{0.4cm}

\begin{center}
{\large\bf Abstract}
\end{center}

We discuss the experimental and theoretical uncertainties on
precision electroweak observables and their relationship to the
indirect constraints on the Higgs boson mass, $\MH$, in the
Standard Model (SM). The critical experimental measurements
($\MW$, $\sweff$, $m_t$, ...) are evaluated in terms of
their present uncertainties and their prospects for improved
precision at future colliders, and their contribution to the
constraints on $\MH$.  In addition, the current uncertainties of
the theoretical predictions for $\MW$ and $\sweff$ due to missing
higher order corrections are  estimated and expectations and
necessary theoretical improvements for future colliders are
explored. The constraints from rare B decays are also discussed.
Analysis of the present experimental and theoretical precisions
yield a current upper bound on $\MH$ of $\sim 200$~GeV.
Including anticipated improvements corresponding to the
prospective situation at future colliders (Tevatron Run~II, LHC,
LC/GigaZ), we find a relative precision of about 25\% to 8\% (or better)
is achievable in the indirect determination of~$\MH$.


\vfill

\begin{picture}(200,0.2)
\put(0,0){\line(1,0){200}}
\end{picture}

$^{\ddag}$ Report of the Precision Electoweak Working Group (P1WG1),
``Workshop on the Future of Particle Physics'', Snowmass, Colorado,
USA, July 2001.

\vspace{0.5em}
$^*$ Working Group conveners




\newpage

\bibliographystyle{revtex}

\title{Present and Future Electroweak Precision Measurements and the
Indirect Determination of the Mass of the Higgs Boson}

\author{{\bf The Precision Electroweak Working Group of Snowmass 2001\\}
U.~Baur}
\affiliation{State University of New York at Buffalo, Buffalo, NY 14260,
USA}

\author{R.~Clare%
\footnote[1]{Working group convener}%
}
\affiliation{University of California, Riverside, CA 92521, USA}

\author{A.~Denner}
\affiliation{Paul Scherrer Institut, Villigen PSI, Switzerland}

\author{J.L.~Diaz~Cruz}
\affiliation{IPUAP-BUAP, 72570 Puebla, Pue., Mexico}

\author{S.~Dittmaier}
\affiliation{DESY, Hamburg, Germany}

\author{J.~Erler%
\footnotemark[1]%
}
\affiliation{Dept.\ of Physics and Astronomy, University of Pennsylvania,
PA 19146, USA and  Universidad Nacional Autonoma de Mexico,Instituto de
Fisica, 01000 Mexico D.F., Mexico}

\author{M.~Gr\"unewald}
\affiliation{RWTH Aachen, Aachen, Germany}

\author{S.~Heinemeyer%
\footnotemark[1]%
}
\affiliation{HET Physics Dept., Brookhaven Natl.\ Lab., NY 11973, USA}

\author{U.~Heintz}
\affiliation{Boston University, Boston, MA 02215, USA}

\author{M.~Kraemer}
\affiliation{University of Edingburgh, Edingburgh, UK}

\author{H.E.~Logan}
\affiliation{Fermilab, PO Box 500, Batavia, IL 60510-0500, USA}

\author{K.~M{\"o}nig}
\affiliation{DESY, Zeuthen, Germany}

\author{M.~Narain}
\affiliation{Boston University, Boston, MA 02215, USA}

\author{M.~Roth}
\affiliation{Universitaet Karlsruhe, Karlsruhe, Germany}

\author{M.~Schmitt}
\affiliation{Northwestern University, Evanston, IL, USA}

\author{D.~Wackeroth}
\affiliation{University of Rochester, Rochester, NY 14627, USA}

\author{G.~Weiglein}
\affiliation{Institute for Particle Physics Phenomenology, Durham, UK}

\author{D.R.~Wood%
\footnotemark[1]%
}
\affiliation{Northeastern University, Boston, MA 02115, USA}

\author{J.~Wudka}
\affiliation{University of California, Riverside, CA 92521}

\date{\today}

\begin{abstract}
We discuss the experimental and theoretical uncertainties on
precision electroweak observables and their relationship to the
indirect constraints on the Higgs-boson mass, $\MH$, in the
Standard Model (SM). The critical experimental measurements
($\MW$, $\sweff$, $\mt$, ...) are evaluated in terms of
their present uncertainties and their prospects for improved
precision at future colliders, and their contribution to the
constraints on $\MH$.  In addition, the current uncertainties of
the theoretical predictions for $\MW$ and $\sweff$ due to missing
higher order corrections are  estimated and expectations and
necessary theoretical improvements for future colliders are
explored. The constraints from rare B decays are also discussed.
Analysis of the present experimental and theoretical precisions
yield a current upper bound on $\MH$ of $\sim 200$~GeV.
Including anticipated improvements corresponding to the
prospective situation at future colliders (Tevatron Run~II, LHC,
LC/GigaZ), we find a relative precision of about 25\% to 8\% (or better)
is achievable in the indirect determination of~$\MH$.
\end{abstract}

\maketitle


\section{Introduction}
\label{sec:mhindirect}

In this contribution we address the status and possible future
developments in the measurements of and the theoretical predictions for
the most important electroweak precision observables. We estimate
their precision from upcoming and proposed accelerator experiments.
In all cases we quote uncertainties which we believe to be realistically
achievable, not excluding even greater precisions.  As a result of
imposing similar standards in all cases, our quoted uncertainties
should be directly comparable.  Similarly, we attempt to anticipate which
improvements can be expected in the theoretical predictions for
the observables.

Within the SM, the mass of the Higgs boson, $M_H$, can be constrained
indirectly
with the help of electroweak precision observables (EWPO).
Among the experimental measurements of EWPO which are used in
global fits, the $W$-boson mass, $\MW$, and the effective leptonic
weak mixing angle, $\sweff$, have the largest impact on the
extracted value of $\MH$. Although the current relative precision
of $\MW$ is better by a factor of 1.8~compared to $\sweff$, the
latter is the most relevant parameter for the indirect $\MH$
determination due to its more pronounced dependence on the Higgs
mass. For equal relative experimental precisions, it yields a
3.1~times higher sensitivity (for $\MH$ around 115~GeV). Other
observables include the leptonic $Z$-boson width, $\gammal$; the
mass and width of the $Z$~boson, $\MZ$ and $\Gamma_Z$; the peak
hadronic cross section of the $Z$~boson, $\sigma^0_{\rm had}$;
EWPO from deep inelastic neutrino scattering; and others.
Furthermore the top quark mass, $\mt$, enters in the global fit;
its value and its error have a strong impact on the extracted
$\MH$ value.  The results from rare B decays do not enter the fit
directly, but they are important for constraining certain non-SM
extensions of the Higgs sector.

    In this article, we begin by summarizing the performance parameters we
have assumed for future colliders.  We then discuss several
critical measurements and the anticipated experimental and
theoretical uncertainties.  Finally, we consider the precision of
future indirect determination of the Higgs-boson mass from
precision electroweak fits. A related discussion of these fits can
be found in Ref.~\cite{Baur:2001yp}.

\section{Collider Parameters}
\label{sec:colliders}

At the time of the present study, LEP has finished running, and nearly
all of the relevant analyses are complete.  The final results from
Tevatron Run~I are all available, and no results are yet available
from Tevatron Run~II.  The b-factories at SLAC and KEK are running
and have some results already available.

A proton-antiproton center-of-mass energy of 2 TeV is assumed for Run~II
of the Tevatron at Fermilab.
Run~IIA is expected to deliver an integrated
luminosity of 2 fb$^{-1}$ to each of the two experiments.  The
next phase of operations, Run~IIB, is expected to deliver 15
fb$^{-1}$ according to the plan currently adopted by Fermilab, but
this could go as high as 30 fb$^{-1}$ if additional upgrades are
made to the accelerator complex.  We refer to these two scenarios
as Run~IIB and Run~IIB*.

The Large Hadron Collider is assumed to deliver 100 \ifb\ to each
experiment each year, with a proton-proton center-of-mass energy
of 14 TeV.

For a future Linear Collider (LC), we assume an electron-positron
(or electron-electron) center-of-mass energy of 0.5 TeV and 500
\ifb\ delivered to a single experiment.
For those measurements which require polarized beams, we make this
assumption explicit.  ``GigaZ'' collectively
denotes an LC operating at $\sqrt{s}=M_Z$ or $\sqrt{s}\approx 2M_W$
with a luminosity of ${\cal L}\approx 5\times 10^{33}{\rm cm^{-2}s^{-1}}$.

\section{The Mass of the $W$ Boson}
\label{sec:mw}

We begin by describing the experimental uncertainties in the
measurement of the $W$-boson mass, first at hadron colliders,
then at lepton colliders.  We then discuss the
theoretical issues.  In the theoretical discussion, we use the
phrase {\em intrinsic\/} uncertainties for the ones arising
from unknown higher-order corrections in the perturbative expansion,
as well as for other uncertainties arising from
other computational limitations.
These are to be distinguished from {\em primordial\/}
uncertainties, which are those uncertainties in relating a
primordial quantity (like the actual mass of the $W$ boson) to the
observables that experiments use to estimate the primordial
quantity.  More details can be found in Ref.~\cite{Baur:2001yp}.

\subsection{Experimental Prospects}

The current precision of $\MW$ is dominated by the direct mass
reconstruction of $W$-pair events at
LEP2. Transverse-mass fits from
Run~I at the Tevatron and data from threshold scans at LEP2 also
contribute significantly but carry less statistical weight.

\smallskip
\noindent\ul{Hadron Colliders:}

The next opportunity to improve the measurement of $\MW$ will be
at Run~II of the Tevatron.  The measurement of the $W$-boson mass
from Tevatron Run~I data achieved a precision of 68
MeV\cite{cdfwmass2,Abazov:2001wq}. A variety of methods can be used to measure the
$W$-boson mass with different tradeoffs between statistical and
systematic uncertainties. These include fits of the transverse
mass and lepton $p_T$ spectra to templates from Monte Carlo
simulations. Most systematics, such as the detector calibration
and the recoil model, are driven by the number of $Z$-boson decays
observed\cite{Abazov:2001wq,cdfwmass2}. Measurements of the charge asymmetry in
$W\to\ell\nu$ decays will help constrain the parton distribution
functions.  In the ratio method, developed by
D\O\cite{D0_Wmass_ratio}, the $Z$-boson data are rescaled to fit
the $W$-boson data. This reduces most experimental and some
theoretical uncertainties at the cost of statistical sensitivity.
In all cases, systematic uncertainties are expected to dominate.
Since the main systematics differ, these methods can be used to
check the results for consistency at the 10 MeV level.

Table \ref{P1WG1_schmitt_0713tab1} shows the expected precision of
the $W$ mass measurement from the transverse mass fit,
extrapolated from the Run~Ib measurement by
D\O\cite{Abbott:1998ww}. The calorimeter scale and linearity
assume constraints from $Z$ data only, but not the $J/\psi$ and
$\pi^0$ data used in Run~I. By about 30 fb$^{-1}$, the
determination of the energy resolution will be systematically
limited by the uncertainty in the width of the $Z$ boson. The
uncertainty due to electron removal was conservatively assumed to
decrease only by half. Table \ref{P1WG1_schmitt_0713tab1} also
shows an extrapolation of the uncertainty for the ratio method
from Run~I results by D\O. The systematic uncertainty for this
method is smaller than for the transverse mass fit and it may well
be the best for high integrated luminosities. We conclude that the
$W$-boson mass will be measured at the end of Run~II to a precision
of 15 MeV, perhaps even 10 MeV, combining the results from both
experiments, using several methods and the $W\to e\nu$ and
$W\to\mu\nu$ channels.

\begin{table}[htbp]
\caption{Projected uncertainties in MeV of the $W$-boson mass
measurement using the transverse mass fit (left) and the ratio
method (right) for $W\to e\nu$ decays.}
\label{P1WG1_schmitt_0713tab1}
\centerline{\hbox to 3in{
\begin{tabular}{l|r|rrr}
 Run & I & IIA & IIB & IIB* \\
$\int{\cal L}dt$ (fb$^{-1}$) & 0.08 & 2 & 15 & 30 \\
\hline\hline
statistical uncertainty & 96 & 19 & 7 & 5 \\
\hline\hline
production/decay model  & 30 & 14 & 13 & 13 \\
backgrounds         & 11 & 2 & 1 & 1 \\
detector model      & 57 & 13 & 8 & 8 \\
\hline
total systematic    & 66 & 19 & 16 & 15 \\
\hline\hline
total uncertainty   & 116 & 27 & 17 & 16 \\
\end{tabular}
}\hbox to 3in{
\begin{tabular}{l|r|rrr}
 Run & I & IIA & IIB & IIB* \\
$\int{\cal L}dt$ (fb$^{-1}$) & 0.08 & 2 & 15 & 30 \\
\hline\hline
statistical uncertainty & 211 & 44 & 16 & 11 \\
\hline
total systematic    & 50 & 10 & 4 & 3 \\
\hline\hline
total uncertainty   & 217 & 44 & 16 & 12 \\
\end{tabular}
}}
\end{table}

The improvement in the measurement of $\MW$ at the LHC is due to the
large statistics which is expected to result in very small statistical
errors and good control of many systematic uncertainties. However, as
in Run~IIB, theoretical improvements are needed, e.g.\ for radiative
$W$ decays, the modeling of the $p_T^W$ distribution, and for
constraining PDFs.  In Ref.~\cite{haywood} it has been argued that it
should be possible to obtain an uncertainty on the $W$ mass due to
PDFs smaller than 10 MeV.

\smallskip
\noindent\ul{Linear Collider:}

As for the $\MW$ measurement
at LEP2, the determination of the $W$ mass at the LC at center of mass (CM)
energies above the $W^+W^-$ production threshold will be based on
direct reconstruction of
$W$-pair events in 4-fermion production processes.
For an integrated luminosity of 500 \ifb\, the uncertainty on
$\MW$ is expected to be about 10 MeV\cite{klausyboy}.

A determination of $\MW$ with the GigaZ option is
based on a dedicated threshold scan.  About one year of
running would be needed to achieve a 5 MeV experimental
error on $\MW$~\cite{wilson}.  This also requires that
 the knowledge of the absolute beam energy can be
controlled to better than 2.5~MeV.
Although this is of higher precision than currently foreseen for TESLA
or NLC, it might be achievable with some additional effort~\cite{wilson}.

\subsection{Theoretical Issues}

For the envisioned precision measurement of the $W$-boson mass, $M_W$,
at present and future lepton and hadron colliders it is crucial that
the theoretical predictions for the underlying production processes
are well under control. The status and prospects of theoretical
predictions for weak gauge boson production processes are presented in
Ref.~\cite{Baur} and briefly summarized in the following.

\smallskip
\noindent\ul{Hadron Colliders}

QED corrections are known to produce a
considerable shift in the $W$- and $Z$-boson
masses measured at hadron
colliders~\cite{cdfwmass,d0wmass,cdfwmass2,d0wmass2}. Given the expected
accuracy for $M_W$ in Run~II of the Tevatron and at the LHC,
calculations including the full ${\cal O}(\alpha)$ electroweak
corrections to weak-boson production in hadronic collisions are
needed.  A calculation of the electroweak ${\cal O}(\alpha)$
corrections to $W$ production has been carried out in Ref.~\cite{BKW}
in the leading-pole approximation, i.e.\ corrections which are very
small at the $W$ pole were ignored. Calculations of the full ${\cal
O}(\alpha)$ corrections to $p\,p\hskip-7pt\hbox{$^{^{(\!-\!)}}$}
\rightarrow W\to\ell\nu_\ell$ have recently appeared in
Refs.~\cite{eww2} and~\cite{Zykunov:2001mn}.
While the corrections ignored in
Ref.~\cite{BKW} change the differential cross section in the $W$-pole
region by less than 1\%~\cite{eww2}, they become large at high
$\ell\nu_\ell$ invariant masses $m(\ell\nu_\ell)$ due to Sudakov-like
logarithms of the form $\ln^2[m(\ell\nu_\ell)/M_W]$.  They
significantly affect the transverse mass distribution above the $W$
peak, which serves as a tool for a direct measurement of the $W$
width, $\Gamma_W$. Taking these corrections into account in future
measurements of $\Gamma_W$ will be important.

The determination of the $W$-boson mass in a hadron collider
environment requires a simultaneous precision measurement of the
$Z$-boson mass, $M_Z$, and width, $\Gamma_Z$. When compared to the
value measured at LEP1, the two quantities help to accurately
calibrate detector
components~\cite{cdfwmass,d0wmass,cdfwmass2,d0wmass2}. It is therefore
necessary to also understand electroweak corrections for $Z$-boson
production.  A calculation of the full ${\cal O}(\alpha)$ QED
corrections to $p\,p\hskip-7pt\hbox{$^{^{(\!-\!)}}$} \to
\gamma,\,Z\to\ell^+\ell^-$ was carried out in Ref.~\cite{BKS}. Purely
weak corrections were ignored in this first step towards a complete
calculation of the ${\cal O}(\alpha)$ electroweak corrections to
$p\,p\hskip-7pt\hbox{$^{^{(\!-\!)}}$} \to \gamma,\,Z\to\ell^+\ell^-$.
However, in order to properly calibrate the $Z$-boson mass and width
using the available LEP data, it is desirable to use exactly the same
theoretical input that has been used to extract $M_Z$ and $\Gamma_Z$
at LEP, i.e.\ to include the purely weak corrections to
$p\,p\hskip-7pt\hbox{$^{^{(\!-\!)}}$} \to \gamma,\,Z\to\ell^+\ell^-$
and the ${\cal O}(G_F^2m^2_tM_W^2)$ corrections to the effective
leptonic weak mixing parameter, $\sweff$, and
the $W$-boson mass~\cite{dg}, in the calculation.
Such a calculation has recently been completed~\cite{ewz2}.
The additional corrections taken into account
in Ref.~\cite{ewz2} enhance the differential cross section in the
$Z$-peak region by up to 1.2\%. Since they are not uniform, they are
expected to shift the $Z$-boson mass extracted from data upward by
several MeV. Detailed simulations of this effect, however, have not
been carried out yet.

For the analysis of Run~IIA data~\cite{Brock:1999ep} the presently
available calculations of $W$- and $Z$-boson production will likely be
sufficient.  However, to measure $M_W$ with a precision of less than
20~MeV in a hadron collider environment as foreseen in Run~IIB and at
the LHC, it will be necessary to take into account higher-order
corrections, in particular multi-photon radiation effects.

\smallskip
\noindent\ul{Lepton Colliders}

At LEP2 and a future Linear Collider (LC), $M_W$ is measured in
$W$-pair production either in a dedicated threshold scan operating the
machine at center-of-mass (CM) energies of $\approx 161$~GeV, or via
direct reconstruction of the $W$ bosons at CM energies above 170 GeV.
For state-of-the-art predictions for $e^+e^-\to WW\to 4f$ cross
sections including ${\cal O}(\alpha)$ corrections, two MC generators
are presently available, {\tt
RacoonWW}~\cite{Denner:1999kn,Denner:1999dt,Denner:2000bj} and {\tt
YFSWW3}~\cite{Jadach:1998hi,Jadach:2000tz,Jadach:2001uu}.  As it is
the case for all present calculations of ${\cal O}(\alpha)$
corrections to $e^+e^-\to WW \to 4f$, they rely on a double-pole
approximation (DPA): electroweak ${\cal O}(\alpha)$ corrections are
only considered for the terms that are enhanced by two resonant $W$
bosons. A tuned numerical
comparison between {\tt RacoonWW} and {\tt YFSWW3}, supported by a
comparison with a semi-analytical calculation~\cite{Beenakker:1999gr}
and a study of the intrinsic DPA ambiguity with {\tt
RacoonWW}~\cite{mclep2,Denner:2000bj} and {\tt YFSWW3}~\cite{mclep2},
shows that the current theoretical uncertainty for the total $W$-pair
production cross section, $\sigma_{WW}$, is about 0.5\% for CM
energies between 170~GeV and 500~GeV~\cite{mclep2}.  Taking the
observed differences~\cite{mclep2} between the {\tt RacoonWW} and {\tt
YFSWW3} predictions at CM energies of 200 and 500 GeV as a guideline,
a theoretical uncertainty of about $1\%$ can be assigned to the
distribution of the $W$ production angle and the $W$ invariant-mass
distribution in the $W$ resonance region.  The theoretical
uncertainties of the $e^+e^-\to WW\to 4f$ cross sections translate
into uncertainties of the $W$ mass and the triple gauge-boson
couplings (TGCs) extracted from data.  A recent
study~\cite{Jadach:2001cz} based on the MC generators {\tt KoralW} and
{\tt YFSWW3} finds a theoretical uncertainty of $\delta M_W=5$~MeV due
to unknown electroweak corrections at LEP2 energies.  Using the MC
generator {\tt YFSWW3}, the ALEPH collaboration~\cite{aleph} has
derived (preliminary) results for the shifts in the extracted values
for the TGCs due to the inclusion of electroweak corrections.
For $M_W$ measurements well above the $W$-pair threshold at LEP2
energies and up to about 500 GeV the accuracy of the corresponding
predictions is sufficient.  More studies, however, are needed to
properly estimate the uncertainties due to missing higher-order
corrections at LC energies above 500 GeV.

In order to meet the goal of a combined error of
$\delta M_W=7$~MeV~\cite{wilson,Baur:2001yp} in a
threshold scan at a LC, the theoretical uncertainty of the
extracted value of $M_W$ has to be less than 2--3 MeV.  At present
there is no study available on how this requirement translates
into a constraint on the theoretical precision for $\sigma_{WW}$
in the threshold region.  The DPA is not a valid approximation in
the threshold region, as the $e^+e^-\to 4f$ cross section in this
region is not dominated by $W$-pair production.  Thus,
$\sigma_{WW}$ in the threshold region is known with an accuracy of
only about 1--2\%~\cite{Beenakker:1996kt,Denner:2001zp}, since
predictions are based on an improved-Born approximation which
neglects non-universal electroweak corrections.  Assuming the
theoretical predictions do not improve, the uncertainty of the $W$
mass obtained from a threshold scan is completely dominated by the
theoretical error, and the precision of the $W$ mass is
limited~\cite{Baur} to about $20-30$ MeV.  Only the knowledge of
the full ${\cal O}(\alpha)$ electroweak corrections to $e^+e^-\to
4f$ in the threshold region will allow one to decide whether the
theoretical uncertainty of the $M_W$ measurement from a threshold
scan can be reduced to the desired level. A full calculation of
the ${\cal O}(\alpha)$ electroweak corrections to $e^+e^-\to WW\to
4f$ is currently not available, but there is ongoing work
in this direction~\cite{Vicini:1998iy,Vicini:2001pd}.  This calculation is of enormous
complexity and the requirement to include finite $W$-width effects
poses severe problems with gauge invariance.

\smallskip
\noindent\ul{Higher Order Uncertainties:}

Concerning the intrinsic uncertainties of $\MW$ from unknown higher
orders, recent progress has been made for the prediction of $\MW$ by
the inclusion of the full fermionic two-loop
corrections~\cite{weiglein}, superseding the previous expansions in
$\mt^2/\MW^2$. Since this expansion yielded similar values (with the
same sign) for the $\mt^4/\MW^4$ and the $\mt^2/\MW^2$ terms (casting
some doubt on the convergence), the full fermionic two-loop corrections
constitute an important step towards a very precise $\MW$ prediction.
The difference between the expansion calculation and the full result can
reach up to about 4~MeV, depending on $\MH$. The only missing two-loop
corrections to $\MW$ are the pure
bosonic contributions. The $\MH$ dependence of the bosonic two-loop
contributions to $\MW$ has recently been evaluated~\cite{MWMHdep2},
indicating corrections of \order{1~{\rm MeV}}.

In order to quantify the remaining intrinsic uncertainties of the EWPO,
one has to perform estimates of the possible size of uncalculated
higher-order corrections.
The results of calculations based on different renormalization
schemes or on different prescriptions for incorporating non-leading
contributions in resummed or expanded form differ from each other
by higher-order corrections. One way of estimating the size of unknown
higher-order corrections is thus to compare the results for the
prediction of the EWPO from different codes in which the same
corrections are organized in a somewhat different way. A detailed
description of different ``options'' used in this comparison can be
found in Ref.~\cite{yellowbook1995} and an update in Ref.~\cite{bardin1999}.
This prescription may lead to an underestimate of the theoretical
uncertainty if at an uncalculated order a new source of potentially
large corrections (e.g.\ a certain enhancement factor) sets in. In
general, it is not easy to quantify how large the variety of different
codes and different ``options'' should be in order to obtain a
reasonable estimate of the theoretical uncertainty.

Other (related) methods to estimate the size of missing higher
order corrections are to vary the renormalization scales and
schemes. While these methods usually give an order of magnitude
estimate and a lower bound on the uncertainty, they can lead to
underestimates whenever there are sizeable but scheme- and
scale-invariant contributions.  For example, the lowest order
flavor singlet contribution to $Z$ decay, a separately gauge
invariant and finite set of corrections, cannot be estimated by
scale variations of the non-singlet contribution or by using
different ``options'' for resumming non-leading contributions in
computer codes.

In the following we use a simple minded, but rather robust and, in the
past, quite successful method for estimating the uncertainties from
unknown higher orders~\cite{GAPP}.
The idea is to collect all relevant enhancement and suppression
factors and setting the remaining coefficient (from the actual loop
integrals) to unity. If, in a given order, terms with different group
theory factors contribute, one can often choose the largest one as
an estimate for the uncertainty.
Our results are summarized in Table~\ref{higherordermw}~\cite{Baur:2001yp}.
They are in good agreement with the estimates of the current
uncertainties of $\MW$ performed in
Refs.~\cite{gambunc,sirlunc,MWMHdep2,freitunc}.

\begin{table}[thb]
\caption{
Theoretical uncertainties from unknown higher-order corrections to
 $\MW$. $\hat{s}$ denotes the \msbar\ mixing angle,
$N=12$ is the number of  fermion doublets in the SM, $C_F = 4/3$ and
$C_A = 3$ are QCD factors, and  $N_C = 3$ is the number of colors.
The corrections in the upper part of the table are assumed to enter the
predictions in the same way as $\Delta\alpha$ (only the leading top quark
correction of \order{\al\als^2} enters via $\De\rho$), while the ones in the
lower part are assumed to enter via $\Delta\rho$.
The fermionic contributions of ${\cal O}(\alpha^2)$ refer to the
non-leading terms beyond the next-to-leading term of the expansion in
powers of $m_t^2/\MW^2$.
The ${\cal O}(\al\als^2)$ corrections,
which are completely known both for $\MW$ and
$\sweff$, are included in the table for completeness.
However, the light fermion corrections are not yet included in all codes
currently used for performing electroweak fits (and have not been
published yet as an independent explicit formula);
our corresponding error estimate for $\MW$ would be
$\pm 1.7$~MeV.
In order to estimate effects of finite $\MH$ and subleading terms
in the lower part of the table, we have taken the average of the
individual coefficients of the result in the limit
$\MH = 0$~\cite{Chetyrkin3} (which in this limit conspire to yield a
small answer), resulting in the numerical prefactors there.
}
\vspace{10pt}
\label{higherordermw}
\renewcommand{\arraystretch}{1.5}
\begin{tabular}{|c|c|c|c|c|}
\hline
 order & sector & estimate & size ($\times 10^{5}$) & $\MW$[MeV]
 \\
\hline \hline
$\al^2$ & fermionic & $N(\al/4\pi\hat{s}^2)^2$ & 8.7
  & complete~\cite{weiglein}  \\
$\al^2$ & bosonic & $(\al/\pi\hat{s}^2)^2$ & 11.6
  & 2.1 \\
$\al\als^2$ & top-bottom doublet & $N_C C_F C_A \al\als^2/4\pi^3\hat{s}^2$
  & 4.7 & complete~\cite{chetyrkin}  \\
$\al\als^2$ & light doublets & $2\; N_C C_F C_A
\al\als^2/4\pi^3\hat{s}^2$ & 9.4 & complete~\cite{chetstein}  \\
\hline
$\al^3\mt^6$ & heavy top & $5.3\; N_C^2(\al\mt^2/4\pi\hat{s}^2\MW^2)^3$
  & 7.0 & 4.1  \\
$\al^3\mt^6$ & heavy top & $3.3\; N_C(\al\mt^2/4\pi\hat{s}^2\MW^2)^3$
  & 1.5 & 0.9 \\
$\al^2\als\mt^4$ & heavy top
  & $3.9\; N_C C_F\al^2\als\mt^4/16\pi^3\hat{s}^4\MW^4$ & 7.8 & 4.5 \\
$\al\als^3\mt^2$ & heavy top &
 $N_C C_F C_A^2 \al\als^3\mt^2/4\pi^4\hat{s}^2\MW^2$ & 2.3 & 1.3
 \\
\hline\hline
& total & & & 7 \\
\hline
\end{tabular}
\renewcommand{\arraystretch}{1}
\end{table}

\section{The effective leptonic weak mixing angle}
\label{sec:sin2th}

\subsection{Experimental Prospects}

The current measurement of $\sweff$ is dominated by the left-right
asymmetry from SLD\cite{Abe:2000dq} and the $b$-quark forward-backward
asymmetry from
LEP1\cite{lepewwg}.

At the Tevatron, the forward-backward asymmetry $A_{FB}$ in the
process $u\overline u+d\overline d\to Z\to \ell^+\ell^-$, measured
near the $Z$ pole, gives a value of the weak mixing angle
$\sweff$. CDF published
$A_{FB}=0.070\pm0.015(\mbox{stat})\pm0.004(\mbox{syst})$ based on
data from Run~I\cite{Abe:1996us}. The statistical uncertainty
scales to 0.0016 for 10 fb$^{-1}$ and to 0.0009 for 30 fb$^{-1}$.
The most important systematic uncertainty arises from the parton
distribution functions. These can be constrained by the charge
asymmetry in $W$ decays. A theoretical uncertainty arises from the
limited rapidity coverage and the $p_T$ distribution of the $Z$.
Both the rapidity and $p_T$ distribution of the $Z$ will be
measured and this uncertainty will be reduced. It is expected that
the statistical uncertainty will dominate all
systematics\cite{Baur:2000cz,Brock:1999ep}.

Combining the electron and muon channels from both CDF and D\O\
and using a linear approximation for the relation between $A_{FB}$
and $\sweff$ leads to a projection for the precision of the
$\sweff$ measurement of 0.00028 for 10
fb$^{-1}$\cite{Baur:2000cz,Brock:1999ep}, to be compared with
the current world-average precision of 0.00017. The precision is
degraded somewhat when the full tree level relation between
$A_{FB}$ and $\sweff$ is taken into account: 0.00029 for 15
fb$^{-1}$ and 0.00020 for 30 fb$^{-1}$.  This is comparable to the
current world average\cite{pdg}, and  should clarify the
3.5$\sigma$ discrepancy between $\sweff$ from $A_{\rm LR}$
measured at SLD\cite{Abe:2000dq} and $A_{FB}^{0,b}$, measured at
LEP\cite{lepewwg}.

In $pp$ collisions at LHC, a forward-backward lepton asymmetry
can also be measured, where the quark direction in the initial
state has to be extracted from the boost direction of the
$\ell^+\ell^-$ system with respect to the beam axis.
In order to improve the experimental
uncertainty of $\sweff$ at the LHC, it will be necessary to detect one
of the leptons originating from $Z\to \ell^+\ell^-$ over the entire pseudorapidity
range of $|\eta|<5$~\cite{haywood}. This requires an electron jet
rejection factor of $<0.01$ in the forward region ($2.5<|\eta|<5$) of the
electromagnetic calorimeter. The relevance of a more precise
determination of PDFs in this respect remains to be investigated.

The determination of the effective leptonic mixing angle at the LC
could in principle be performed by a fixed-target M\o ller
scattering experiment~\cite{kumar,baurdem} (but see also
Ref.~\cite{gambinocuypers}.)  However, this would yield
$\sweff$ at a scale of \order{0.5 {\rm ~GeV}}, making the
extrapolation to $\MZ$ difficult. We therefore do not
consider this option here in more detail (although sensitivity to physics
beyond the SM might be possible.)

At GigaZ one hopes to improve the current precision of $\sweff$ by
more than an order of magnitude.
This is envisaged by a precise measurement of
$A_{\rm LR}$~\cite{moenig,sn_pol} using the Blondel scheme~\cite{blondel}.
$A_{\rm LR}$ is then given as a function of polarized cross sections,
where both beams have possibly different combinations of polarizations.
Due to the anticipated drastic improvement in the accuracy, a
reanalysis of the effect of primordial uncertainties in the
determination of $\sweff$ might become necessary.
This determination of $A_{\rm LR}$ requires that both beams can be
polarized independently and that
the polarizations of the colliding $e^+$~and $e^-$~bunches with
opposite helicity states are equal (or that their difference is
precisely determined; see Ref.~\cite{moenig} for details). A precision of
$\de A_{\rm LR} \approx 8 \times 10^{-5}$ seems to be
feasible~\cite{moenig,teslatdr}, resulting in
$\de\sweff \approx 10^{-5}$.

\subsection{Theoretical Issues}

Concerning the intrinsic uncertainties of $\sweff$ from unknown higher
orders,  the
situation is slightly worse than for $\MW$, since a result
for the full fermionic
two-loop corrections is not yet available, and one has to rely on the
expansion in powers of $m^2_t/\MW^2$~\cite{sweffexpansion}.
Beyond two-loop order, the results for the pure fermion-loop
contributions (incorporating in particular the leading terms in
$\Delta\alpha$ and $\Delta\rho$) are known up to the four-loop
order~\cite{fermloop}. Furthermore, the QCD corrections of
${\cal O}(\al\als^2)$ are known~\cite{chetyrkin,chetstein}. More
recently, also the leading three-loop terms
of ${\cal O}(G_F^3 m_t^6)$ and ${\cal O}(G_F^2 \als m_t^4)$, which enter
via the quantity $\Delta\rho$, have been calculated in the limit of
vanishing Higgs-boson mass. The results have been found to be quite
small, which is familiar from the $\MH = 0$ limit of the
${\cal O}(G_F^2 m_t^4)$ result~\cite{vanderbij}. In the latter case,
the extension to finite values of $\MH$ and the inclusion of subleading
terms led to an increase in the numerical result by a factor of up to 20.

A shift in the prediction
for $\MW$, on the other hand, induces a shift in $\sweff$
according to

\begin{equation}
\sweff = \left(1 - \frac{\MW^2}{\MZ^2} \right) \kappa(\MW^2) ,
\label{eq:sweffshift}
\end{equation}
where $\kappa$ is a calculable function in the SM.
While the shift in $\MW$ induced by going from the result of the
expansion in powers of $m^2_t/\MW^2$ to the result of the full fermionic
two-loop corrections is known, the corresponding result for
$\kappa(\MW^2)$ is still missing.
The effect of inserting the new result for $\MW$ in
Eq.~(\ref{eq:sweffshift}), which amounts to an upward shift of about
$8 \times 10^{-5}$ in $\sweff$ (for $\MH \approx 115$~GeV), has been
(conservatively) treated as
a theoretical uncertainty in the ``Blue Band'' of Fig.~\ref{blueband}
(see Sec.~\ref{ewfits}).

The same method described earlier for $\MW$ is used to
quantify the remaining intrinsic uncertainties of $\sweff$~\cite{GAPP}.
Our results are summarized in Table~\ref{higherordersweff}, with
a total contribution to the uncertainty of $7 \times 10^{-5}$.
They are in good agreement with the estimates of the current
uncertainties of $\sweff$ performed in
Refs.~\cite{gambunc,sirlunc,MWMHdep2,freitunc}.

\begin{table}[thb]
\caption{
Theoretical uncertainties from unknown higher-order corrections to
$\sweff$.
See Table~\ref{higherordermw} for a key to the notation and
assumptions. The uncertainty in $\sweff$ has been
estimated from the known correction to $\MW$ using
Eq.~(\ref{eq:sweffshift}) (see text).
The light fermion corrections are not yet included in all codes
currently used for performing electroweak fits (and have not been
published yet as an independent explicit formula);
our corresponding error estimate for $\sweff$ would be
 $\pm 3.3\times 10^{-5}$.
}
\vspace{10pt}
\label{higherordersweff}
\renewcommand{\arraystretch}{1.5}
\begin{tabular}{|c|c|c|c|c|}
\hline
 order & sector & estimate & size ($\times 10^{5}$)
  & $\sweff$ ($\times 10^{5}$) \\
\hline \hline
$\al^2$ & fermionic & $N(\al/4\pi\hat{s}^2)^2$ & 8.7
  &  4.1 \\
$\al^2$ & bosonic & $(\al/\pi\hat{s}^2)^2$ & 11.6
 & 4.1 \\
$\al\als^2$ & top-bottom doublet & $N_C C_F C_A \al\als^2/4\pi^3\hat{s}^2$
  & 4.7  & complete~\cite{chetyrkin} \\
$\al\als^2$ & light doublets & $2\; N_C C_F C_A
\al\als^2/4\pi^3\hat{s}^2$ & 9.4 &
complete~\cite{chetstein} \\
\hline
$\al^3\mt^6$ & heavy top & $5.3\; N_C^2(\al\mt^2/4\pi\hat{s}^2\MW^2)^3$
  & 7.0 & 2.3 \\
$\al^3\mt^6$ & heavy top & $3.3\; N_C(\al\mt^2/4\pi\hat{s}^2\MW^2)^3$
  & 1.5 & 0.5 \\
$\al^2\als\mt^4$ & heavy top
  & $3.9\; N_C C_F\al^2\als\mt^4/16\pi^3\hat{s}^4\MW^4$ & 7.8 & 2.5 \\
$\al\als^3\mt^2$ & heavy top &
 $N_C C_F C_A^2 \al\als^3\mt^2/4\pi^4\hat{s}^2\MW^2$ & 2.3 & 0.8
 \\
\hline\hline
& total & &  & 7 \\
\hline
\end{tabular}
\renewcommand{\arraystretch}{1}
\end{table}

\section{Top quark properties}
\label{sec:top}

For the precision EW fits, the most important top quark property
is its mass, $\mt$.  Although from an experimental standpoint,
$\mt$ is on the same footing as $\MW$ and $\sweff$, theoretically
it is handled a bit differently and affects the theoretical
predictions for $\MW$ and $\sweff$.

In Tevatron Run~I, the top quark mass was measured to $\approx5$
GeV\cite{Affolder:2000vy,Abbott:1998dc}. For Run~II, $t\overline t$
data samples will be large enough to allow a double $b$-tag. For
15 fb$^{-1}$ per experiment, 3200 double-tagged single-lepton and
1200 untagged dilepton events are expected.

The main systematic uncertainty for the top quark mass measurement
is the jet energy scale. Using Run~I methods, this uncertainy
cannot be reduced below a couple of GeV. However, both experiments
plan to use $p\overline p\to Z\to b\overline b$ events, which will
help set the energy scale to a precision of about 0.5
GeV\cite{Heintz_Zbb}. In addition, the hadronically decaying $W$
in single-lepton $t\overline t$ events provides an independent
calibration point\cite{Frey:1997sg}.

The next most important systematic uncertainty is modeling of
initial and final state gluon radiation in  $t\overline t$ events.
In Run~I, this uncertainty was estimated mainly by comparing
predictions of different event generators. In Run~II, the modeling
of jet activity in top quark events can be constrained better by
comparing double-tagged events with simulations. We estimate this
uncertainty to be about 1~GeV, and expect it to decrease only
slightly with increasing integrated luminosity. Other systematics
will scale inversely with the square root of the integrated
luminosity.

We extrapolate the uncertainty on the top mass based on Run~I D\O\
results in the single-lepton channel\cite{Abbott:1998dc} in
Table \ref{P1WG1_schmitt_0713tab3}. We take the higher cross
section at $\sqrt{s}=2$ TeV in account and we assume double
$b$-tagging with an efficiency per $b$-jet of 65\%. Double
$b$-tagging will essentially eliminate the uncertainty due to the
$W$+jets background. The uncertainty in the dilepton
channel\cite{Abbott:1998dn} is also extrapolated in Table
\ref{P1WG1_schmitt_0713tab3}. No $b$-tagging is assumed here. For
each channel and each experiment, a precision of abut 1.2 GeV is
projected. By combining both channels and experiments an overall
precision close to 1 GeV should be achievable.

\begin{table}[htbp]
\caption{Projected uncertainties in GeV of the top quark mass
measurement in the single-lepton channel (left) and dilepton
channel (right).}
\label{P1WG1_schmitt_0713tab3}
\centerline{\hbox to 3in{
\begin{tabular}{l|r|rrr}
 Run & I & IIA & IIB & IIB* \\
$\int{\cal L}dt$ (fb$^{-1}$) & 0.1 & 2 & 15 & 30 \\
\hline\hline
statistical uncertainty & 5.6 & 1.7 & 0.63 & 0.44 \\
\hline\hline
jet scale ($W\to q\overline q$) & 4.2 & 1.8 & 0.64 & 0.45 \\
jet scale ($Z\to b\overline b$) & --- & 0.53 & 0.19 & 0.14 \\
MC model (gluon radiation) & 1.9 & 1.1 & 0.97 & 0.96 \\
event pile-up       & 1.6 & 0.49 & 0.18 & 0.13 \\
$W$+jets background & 2.5 & 0 & 0 & 0 \\
$b$-tag         & 0.4 & 0 & 0 & 0 \\
\hline
total systematic    & 5.5 & 2.1 & 1.2 & 1.1 \\
\hline\hline
total uncertainty   & 7.8 & 2.7 & 1.3 & 1.2 \\
\end{tabular}
}\hbox to 3in{
\begin{tabular}{l|r|rrr}
 Run & I & IIA & IIB & IIB* \\
$\int{\cal L}dt$ (fb$^{-1}$) & 0.1 & 2 & 15 & 30 \\
\hline\hline
statistical uncertainty & 12.3 & 2.4 & 0.87 & 0.62 \\
\hline\hline
jet scale       & 2.0 & 0.88 & 0.32 & 0.23 \\
MC model        & 2.3 & 1.0 & 0.96 & 0.96 \\
event pile-up       & 1.4 & 0.27 & 0.10 & 0.07 \\
background      & 0.9 & 0.17 & 0.06 & 0.05 \\
method          & 0.9 & 0.17 & 0.06 & 0.05 \\
\hline
total systematic    & 3.6 & 1.4 & 1.0 & 1.0 \\
\hline\hline
total uncertainty   & 12.8 & 2.8 & 1.3 & 1.2 \\
\end{tabular}
}}
\end{table}

At the LC, the top quark mass can either be extracted from a
$t\bar t$ threshold scan that would determine a suitably defined
threshold mass~\cite{tttheo,ttexp}, or in the continuum by direct
kinematical reconstruction of
$e^+e^-\to\bar tt\to W^+W^-b\bar b\to \ell^+\nu \ell^-\bar\nu b\bar b$
events~\cite{yeh} which determines the pole mass. The remaining
theoretical uncertainties are sufficiently small to allow a
measurement of the threshold mass with a precision of $\sim
100$~MeV~\cite{tttheo,ttexp,mtnew}.
(For the conversion of the \msbar\ top quark mass to other top quark
mass definitions see e.g. Ref.~\cite{Chetyrkin:1999qi,Melnikov:2000qh}
 and references therein.)
The measurement of the pole mass at higher energies
with an accuracy of 200 MeV or better may be possible~\cite{yeh},
but is limited in precision
by QCD renormalon effects which are of ${\cal O}(\Lambda_{QCD})$.

The precise measurement of $m_t$ at the LC will eliminate the
main source of parametric uncertainties of the EWPO. The
uncertainties induced in $\MW$ and $\sweff$ by the experimental
error of $m_t$ will be reduced by the LC measurement to the
level of 1~MeV and $0.5 \times 10^{-5}$, respectively, i.e.\
far below the uncertainty corresponding to the present error of
$\de\De\al_{\rm had}$.

The $t \bar t$ threshold analysis at the LC will result in correlated
measurements of $\als$ and $\mt$. Since an independent
and more precise determination of $\als(\MZ)$ would be possible with GigaZ
(to $\pm 0.0010$, from other GigaZ
observables: the $Z$~width with 1~MeV uncertainty, and $R_l$ with 0.05\%
uncertainty~\cite{jenssvengigaz,alphas}), an
improved value for $\mt$ can be expected as well.

\section{Rare B decays}
\label{sec:rareb}

Rare $B$ meson decay results are not among the inputs to the global fits
for $\MH$, but they provide important constraints on the EW
symmetry breaking sector, so they were included in this study.
Rare $B$ meson decays that first appear at one loop in the
Standard Model (SM) are especially sensitive to new physics that
also enters at one loop.  We considered the processes $b \to s
\gamma$, $B \to \ell^+\ell^-$, and $B \to X_s \ell^+ \ell^-$.

The decay $\bar B \to X_s \gamma$ has been
measured~\cite{bsgexpts} and is in agreement with the SM
prediction.  This allows constraints to be placed on new physics,
most notably the two Higgs doublet model (2HDM)~\cite{2HDMbsg} and
the Minimal Supersymmetric Standard Model (MSSM)~\cite{MSSMbsg}.
The constraints on new physics depend sensitively on the SM
result; recent inclusion of quark mass
effects~\cite{Gambino:2001ew} in the SM calculation shifted the SM
prediction by $\sim 1 \sigma$. The current SM theoretical
uncertainty in BR($\bar B \to X_s \gamma$) is about the same size
as the experimental uncertainty and is limited by the charm quark
mass dependence~\cite{Gambino:2001ew}. The BaBar and BELLE
experiments expect to reduce the current 12\% statistical
uncertainty in BR($\bar B \to X_s \gamma$) to about 5\%, which is
limited by the amount of running off the $\Upsilon(4s)$ resonance
to measure backgrounds.

The decay $B \to \ell^+ \ell^-$ is highly suppressed in the SM and
has not been observed.  The best current bound is BR$(B_s \to
\mu^+ \mu^-) < 2.6 \times 10^{-6}$~\cite{Abe:1998ah} while the
corresponding SM prediction is
$\sim 4 \times 10^{-9}$~\cite{Logan:2000iv,bsmmSM}. New
physics contributions to this decay from the 2HDM can enhance or
suppress the branching ratio by a factor of
two~\cite{Logan:2000iv}, while in the MSSM a much more dramatic
enhancement is possible, even up to the current experimental bound
at large $\tan\beta$ values~\cite{MSSMBll}. Both CDF and D\O\ expect
to observe a few events at the SM rate with 15 fb$^{-1}$ at Run~II
of the Tevatron; severe constraints or a discovery are possible in
the MSSM.

The decay $B \to X_s \ell^+ \ell^-$ is not as highly suppressed
and inclusive branching ratios of several $\times 10^{-6}$ are
expected in the SM; experimental bounds from BaBar are BR$(B \to K
\ell^+\ell^-) < 0.6 \times 10^{-6}$ and BR$(B \to K^*
\ell^+\ell^-) < 2.5 \times 10^{-6}$~\cite{Aubert:2001xt}, close to
the SM predictions. BELLE has recently reported observation of $B
\to K \mu^+ \mu^-$~\cite{BELLEBKmumu}. The spectrum and
asymmetries in this decay allow separate measurements of the signs
and magnitudes of three Wilson coefficients~\cite{Ali:1999mm} and
are sensitive to MSSM contributions. More data is expected soon at
BaBar and BELLE. CDF and D\O\ expect to be able to measure the BR,
spectrum, and energy-dependent asymmetry for $B \to K^* \mu^+
\mu^-$  at Run~II of the Tevatron.  Further refinements in the
asymmetry and spectrum can be expected at BTev.


\section{Electroweak Global Fits}
\label{ewfits}

The LEP Electroweak Working Group's most recent fit to the
current precision electroweak data is shown in
Fig.~\ref{blueband}~\cite{ewwg}.  The figure
shows $\De\chi^2 \equiv \chi^2 - \chi^2_{\rm min}$ as
an approximately quadratic function of $\log M_H$. Therefore, the 95\%
CL upper
limit can be approximated by $\De\chi^2 = 2.71$, corresponding to a
95\% CL upper bound of $M_H < 196$~GeV at present.

\begin{figure}[t]
\includegraphics[height=10cm]{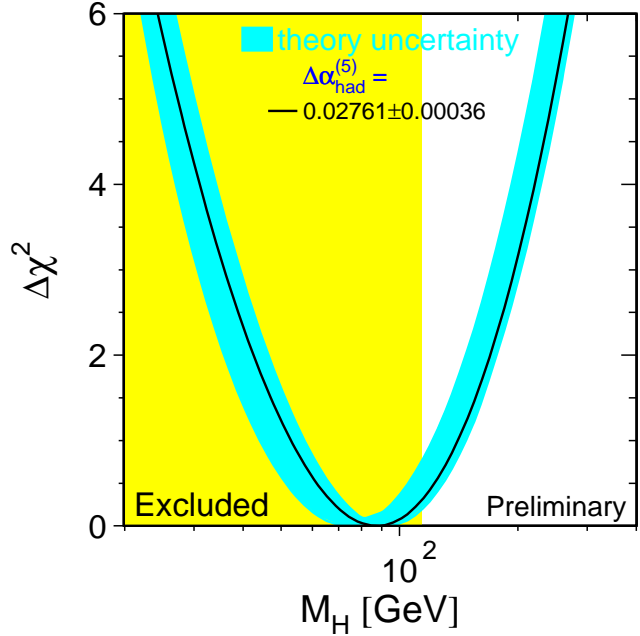}
\caption{$\De\chi^2=\chi^2 - \chi^2_{\rm min}$ from a global fit to all
available data~\cite{ewwg} as a function of the SM Higgs-boson mass,
$M_H$. The width of
the ``Blue Band'' indicates the effect of intrinsic uncertainties
from unknown higher order corrections. The yellow region is
excluded by direct Higgs searches at LEP2~\cite{lephiggs}.}
\label{blueband}
\end{figure}

The precision of the fit results depends on the experimental
uncertainties of the measured values of the EWPO and the
theoretical uncertainties of their predictions.
The SM predictions for the EWPO are calculated in terms of a small set
of input parameters:
$M_Z, G_{\mu}, \alpha (M_Z), m_\ell, m_q,
\mt, \MH$, and $\alpha_s(M_Z)$.  The fine structure constant, $\alpha(0)$,
the $Z$-boson mass, $\MZ$, the lepton masses,
$m_\ell$, and the Fermi constant, $G_F$, are currently the most
precisely measured input parameters~\cite{pdg}, and their errors have
negligible effects on the fit results~\cite{Erler:1995fz,lepewwg,pgl}.
The dominant uncertainties presently arise from the experimental error on
the top quark mass, $\mt = 174.3 \pm 5.1$~GeV~\cite{pdg}, the hadronic
contribution to the fine structure constant at the $Z$-boson mass,
$\De\alpha_{\rm had}$~\cite{jegerlehner,deltaalpha}
(the value used in Fig.~\ref{blueband} is from Ref.~\cite{deltaalpha1}),
as well as $\MH$. $\alpha_s(M_Z)$ is constrained mainly by
$\Gamma_Z$, $R_l$, and $\sigma^0_{\rm had}$, with little theoretical
uncertainty as long as one ignores the possibility of large new physics
effects.

In practice, both EWPO and input parameters are used as
constraints in the fits subject to their experimental
uncertainties (which, as explained above, contain the primordial
theoretical uncertainties related to extraction of the EWPO). The
only distinction is that the input parameters are treated as fit
parameters, and the EWPO are computed in terms of these.  For
example, $m_t$ which appears only in loops is chosen as input.
Moreover, one usually prefers to compute less precise quantities
in terms of more precise ones.  The fit results are insensitive to
these choices.

Table~\ref{expfuture} summarizes the current status of the
experimental uncertainties and the precision one expects to
achieve at future colliders for the most relevant EWPO, $\MW$ and
$\sweff$, and the top quark mass, together with the expected
experimental error on $\MH$\cite{Gunion:1996cn}, assuming the SM Higgs
boson has been discovered with $\MH \approx 115$~GeV. The entries
in the table attempt to represent the combined results of all
detectors and channels at a given collider, taking into account
correlated systematic uncertainties.

\begin{table}[bht]
\caption{
The expected experimental uncertainties (including theory errors for the
experimental extraction)
at various colliders are summarized
for $\sweff$, $\MW$, $\mt$, and $\MH$ (the latter assuming
$\MH = 115$~GeV\cite{p1wg2}).
Each column represents the combined results of all detectors and
channels at a given collider, taking into account correlated
systematic uncertainties.
(The entry in parentheses assumes a fixed target polarized M\o ller scattering
experiment using the $e^-$ beam~\cite{kumar,baurdem}, thus corresponding
to an effective mixing angle at a scale of
${\cal O}(0.5$~GeV). It is not used in the fits.)
}
\vspace{10pt}
\label{expfuture}
\renewcommand{\arraystretch}{1.5}
\begin{tabular}{|c||c||c|c|c|c||c|c|}
\cline{2-8} \multicolumn{1}{c||}{}
& now & Tev.\ Run~IIA & Run~IIB & Run~IIB$^*$ & LHC & ~LC~  & GigaZ \\
\hline\hline
$\de\sweff(\times 10^5)$ & 17   & 78   & 29   & 20   & 14--20 & (6)  & 1.3  \\
\hline
$\de\MW$ [MeV]           & 33   & 27   & 16   & 12   & 15   & 10   & 7      \\
\hline
$\de\mt$ [GeV]           &  5.1 &  2.7 &  1.4 &  1.3 &  1.0 &  0.2 & 0.13   \\
\hline
$\de\MH$ [MeV]            &  --- &  --- &
                 \multicolumn{2}{c|}{${\cal O}(2000)$} &  100 &   50 &   50 \\
\hline
\end{tabular}
\renewcommand{\arraystretch}{1}
\end{table}

The dominant parametric uncertainty of the EWPO presently arises from
the experimental error of the top quark mass, $\de\mt=5.1$~GeV.
This error induces a parametric uncertainty of 32~MeV and
$16 \times 10^{-5}$ in
the $W$~mass and the leptonic effective mixing angle, respectively.
The value of $\De\alpha_{\rm had}$ used in the fit is calculated according
to~\cite{deltaalpha}, with the unsubtracted dispersion relation
(UDR) approach, and the
corresponding errors from its uncertainty,
$\de\De\alpha_{\rm had} = 0.0002$, are 3.7~MeV and
$7 \times 10^{-5}$.  Furthermore, the imperfect knowledge of the
strong coupling
constant, $\de\alpha_s(M_Z) = 0.0028$, introduces uncertainties of 2~MeV
and $3.5 \times 10^{-5}$ and also an uncertainty in $\De\al_{\rm had}$
of about $\de\De\al_{\rm had} = 0.0001$.
While the uncertainty induced by the top quark
mass is about as large as the present experimental error of $\MW$
and $\sweff$, the parametric uncertainties induced by the errors of
$\De\alpha_{\rm had}$ and $\alpha_s(M_Z)$ are already smaller than the
prospective experimental errors on $\MW$ and $\sweff$ at the Tevatron
and the LHC (see Table~\ref{expfuture}). On the other hand, the
accuracies reachable at GigaZ will clearly require a significantly
improved experimental precision not only of $\mt$ (see
Table~\ref{expfuture}), but also of $\De\alpha_{\rm had}$ and
$\alpha_s(M_Z)$. An improved determination of $\alpha_s(M_Z)$ with
little theoretical uncertainty is, in fact, expected from GigaZ
itself~\cite{jenssvengigaz,alphas}.

The ``Blue Band'' shown in Fig.~\ref{blueband} is obtained
by comparing the predictions of the EWPO using the codes {\tt
ZFITTER}~\cite{ZFITTER} and {\tt TOPAZ0}~\cite{TOPAZ0}. At
present, the theoretical uncertainty represented by the width of
the ``Blue Band'' mainly arises from the intrinsic uncertainties
in the prediction for $\sweff$, while the intrinsic uncertainty in
the prediction for $\MW$, being significantly smaller than the
experimental error, is less important.

We have performed a global fit to all data in the Standard Model based
on the present experimental and parametric uncertainties and using the
estimates of Tables \ref{higherordermw} and \ref{higherordersweff}
for the intrinsic theoretical
uncertainties from unknown higher-order corrections. For the
theoretical predictions the program {\tt GAPP}~\cite{GAPP} has been used.
In contrast to the fit in Fig.~\ref{blueband}, where the theory
uncertainties are represented by the width of the blue band,
we have added theoretical and experimental errors in quadrature. As a
result we find
\begin{equation}
\MH = 97^{+ 53}_{- 36} {\rm ~GeV},
\label{eq:presentfit}
\end{equation}
and a 95\% CL upper bound of $\MH < 194$~GeV.
These numbers are very close to the result of the fit in
Fig.~\ref{blueband}~\cite{ewwg}.

Concerning the interpretation of the fit result, it should be kept in
mind that it is based on the assumption that the Standard Model
provides the correct description of the experimental measurements.
This means, in particular, that the resulting bound on $\MH$ does not
reflect the quality of the fit, i.e.\ it does not
contain information about how well the SM actually describes
the data.


\section{Future indirect determinations of $\MH$}
\label{sec:futuremh}

For the analysis in this section, we
anticipate that in the future the currently missing corrections
indicated in the upper part of Tables \ref{higherordermw}
and \ref{higherordersweff} will
become available,
and that the uncertainties listed in the lower part will be reduced
by a factor of two.

In the following we will discuss the anticipated future
experimental precisions of the EWPO reachable at the next
generation of colliders as given in Table~\ref{expfuture} in
view of necessary improvements of the primordial theoretical
uncertainties. In each case we also
investigate whether the prospective parametric and
intrinsic theoretical uncertainties of the EWPO will be sufficiently
under control in order to match the projected experimental precision.

The improvement in the experimental determination of $m_t$ at
Run~IIA will reduce the parametric theoretical uncertainties of
$\MW$ and $\sweff$ to values below the experimental errors of
these observables. Similarly, the present values of the
intrinsic theoretical uncertainties of $\MW$ and $\sweff$ (see
Tables \ref{higherordermw} and \ref{higherordersweff}) are smaller
than the envisaged
experimental errors (see Table~\ref{expfuture}). On the other
hand, an improvement of the theoretical prediction of $\sweff$,
in particular the inclusion of the missing corrections of
${\cal O}(\alpha^2)$, would lead to a significant reduction of
the width of the ``Blue Band'' shown in Fig.~\ref{blueband}. In
the near future, the full two-loop results for $\MW$ and
$\sweff$ constitute the most important contributions in the
MSM theory calculations, and are needed on the time scale of
Run~II, which is taking place over the next $\sim5$ years.

In view of the increased precision of $\sweff$ at GigaZ and the
largely reduced error of $\mt$ at the LC, it will be
very important to reduce the
uncertainty of $\de\De\alpha_{\rm had}$ at least to the level of
$\de\Delta\alpha (M_Z) = 7\times 10^{-5}$~\cite{jegerlehner},
corresponding to parametric uncertainties of $\MW$ and $\sweff$
of $1.5$~MeV and $2.5 \times 10^{-5}$, respectively.
This will require improved measurements of
$R \equiv \sigma(e^+e^- \to {\rm hadrons})/
          \sigma(e^+e^- \to \mu^+\mu^-)$
below about $\sqrt{s} \le 5$~GeV.
In case the uncertainty of $\Delta\alpha (M_Z)$
could even be improved by another factor of two (and taking also
into account the expected improvement in the $\als(M_Z)$
determination at GigaZ), the limiting factor in the parametric
uncertainty of $\sweff$ would arise from the experimental error
of $M_Z$ ($\de M_Z = 2.1$~MeV induces an uncertainty of
$1.4 \times 10^{-5}$ in $\sweff$), which is not expected to
improve in the foreseeable future.

With the prospective future improvements of higher order
corrections to the EWPO discussed above (i.e.\ complete
electroweak two-loop results and a reduction of the
uncertainties in the lower parts of Tables \ref{higherordermw}
and \ref{higherordersweff}
by a factor of two), the intrinsic theoretical uncertainties
of the EWPO will be comparable to or smaller than the parametric
uncertainties and the experimental errors at GigaZ (see also
Ref.~\cite{gigaztheo}.)

In summary, the projected experimental accuracies at
GigaZ require on the theory side a
considerable effort to reduce primordial
theoretical uncertainties. In addition, improvements of the
intrinsic and parametric uncertainties of the EWPO are needed. These tasks
appear challenging, but, in view of the time scale of at least a
decade, not unrealistic.

Based on the uncertainties expected at
the next generation of colliders and our estimates of
present and future theoretical uncertainties, we list in
Table~\ref{indirectmh} the (cumulative) precision of $\MH$ at different
colliders which one hopes to achieve from EWPOs.
Results are given for $\delta \MH/\MH$ obtained from $\MW$
alone, from $\sweff$ alone, and from all precision data, taking into
account the intrinsic and the parametric theoretical uncertainties and
their correlated effects.  These future estimates range from 25\% at the
end of Tevatron Run IIB to 18\% after LHC to 8\% after LC/GigaZ.

\begin{table}[thb]
\caption{The expected {\sl cumulative\/} precision, $\de\MH/\MH$, from future
collider data, given the error projections in
Tables~\ref{expfuture}, \ref{higherordermw},
and~\ref{higherordersweff}. Intrinsic theoretical
and parametric uncertainties and their correlated effects
on $\MW$ and $\sweff$ are taken into account.
In the first row, our estimate for the current intrinsic
uncertainties in $\MW$ and $\sweff$ from unknown higher order
corrections as given in Tables~\ref{higherordermw}
and~\ref{higherordersweff} is used.
In the other rows we assume that complete two-loop results for
the most relevant EWPO are available, and that the
uncertainties in the lower parts of Tables~\ref{higherordermw}
and~\ref{higherordersweff}
have been reduced by a factor of two. This corresponds to
future intrinsic theoretical uncertainties in $\MW$ and $\sweff$
of 3~MeV and $1.7\times 10^{-5}$, respectively$^a$.
As in Eq.~(\ref{eq:presentfit}) we have added the theoretical and
experimental errors in quadrature. We also assume
$\de\Delta\alpha (M_Z) = 7\times 10^{-5}$~\cite{jegerlehner}.
(Using the very optimistic value of $5\times 10^{-5}$ would improve
the $\de\MH$ uncertainty at GigaZ to 7\%.)
The last row also assumes a determination of $\alpha_s (M_Z)$
with an uncertainty of $\pm 0.0010$ from other GigaZ observables. }
\label{indirectmh}
\renewcommand{\arraystretch}{1.5}
\begin{tabular}{|l||r|r||r|}

\hline
$\delta\MH/\MH$ {\it from:}            & $\MW$ & $\sweff$ & all  \\
\hline \hline
now                                    &106 \% & 60 \%    & ~58 \% \\
\hline \hline
Tevatron Run IIA                       & 72 \% & 39 \%    & ~35 \% \\
\hline
Tevatron Run IIB                       & 37 \% & 33 \%    & ~25 \% \\
\hline
Tevatron Run IIB$^*$                   & 30 \% & 29 \%    & ~23 \% \\
\hline
LHC                                    & 22 \% & 25 \%    & ~18 \% \\
\hline \hline
LC                                     & 15 \% & 24 \%    & ~14 \% \\
\hline
GigaZ                                  & 12 \% &  8 \%    &  ~8 \% \\
\hline
\end{tabular}
\renewcommand{\arraystretch}{1}
\end{table}

\renewcommand{\thefootnote}{\alph{footnote}}
\footnotetext[1]{Technically within {\tt GAPP} this is realized by
assuming an uncertainty in the $T$ parameter, $T = 0 \pm 0.007$.}
If the SM is the correct low energy theory, the Higgs boson will be
discovered in Run~II at the Tevatron or at the LHC. In this case, the
indirect determination of $\MH$ from precision electroweak measurements
will constitute an important internal
consistency check of the SM. Possible new scales beyond the SM could
manifest themselves in a disagreement of the directly and
indirectly
determined $\MH$ value~\cite{jenssvengigaz,sitgesproc}.


\begin{acknowledgments}

U.B. is supported by NSF grant PHY-9970703.
The work of D.W. is supported by the U.S. Department
of Energy under grant DE-FG02-91ER40685.
D.R.W. is supported by NSF grant PHY-9972170.
Fermilab is operated by Universities Research Association Inc.\
under contract no.\ DE-AC02-76CH03000 with the U.S.\ Department of
Energy.

\end{acknowledgments}


\bibliography{p1wg1}

\end{document}